# Effect of growth temperature on the CVD grown Fe filled multi-walled carbon nanotubes using a modified photoresist

Joydip Sengupta [a], Avijit Jana [b], N.D. Pradeep Singh [b], Chacko Jacob [a]

[a] Materials Science Centre, Indian Institute of Technology, Kharagpur 721302, India
[b] Department of Chemistry, Indian Institute of Technology, Kharagpur 721302, India

Keywords:
A.   Nanostructures
B.   Vapor deposition
C.   Electron microscopy
C.   Raman spectroscopy
C.   X-ray diffraction

**ABSTRACT**

Fe filled carbon nanotubes were synthesized by atmospheric pressure chemical vapor deposition using a simple mixture of iron(III) acetylacetonate (Fe(acac)$_3$) with a conventional photoresist and the effect of growth temperature (550–950 °C) on Fe filled nanotubes has been studied. Scanning electron microscopy results show that, as the growth temperature increases from 550 to 950 °C, the average diameter of the nanotubes increases while their number density decreases. High resolution transmission electron microscopy along with energy dispersive X-ray investigation shows that the nanotubes have a multi-walled structure with partial Fe filling for all growth temperatures. The graphitic nature of the nanotubes was observed via X-ray diffraction pattern. Raman analysis demonstrates that the degree of graphitization of the carbon nanotubes depends upon the growth temperature.

## 1. Introduction

In today's world of nanotechnology, carbon nanotubes (CNTs) have become a high priority material because of their unique properties [1] with enormous potential in many technological applications. Among the different synthesis methods of CNTs [2–6], chemical vapor deposition (CVD) is the most preferable method owing to its advantage of producing a large amount of CNTs directly on a desired substrate with high purity. Before the synthesis of CNTs by CVD, high temperature hydrogen treatment of the catalyst is required in order to produce contamination-free catalytic particles and for the removal of oxides that may exist over the catalyst surface [7]. In recent years, magnetic material filled CNTs have become an area of interest for the researchers as they have extended the potential applications to magnetic force microscopy [8], high density magnetic recording media [9] and biology [10]. Though different methods of filled CNT synthesis including capillary incursion [11], chemical method [12], arc-discharge [13] and CVD [14–17] have already been reported, yet the processes are costly as they usually require a two-stage process along with rigorous control of the reaction parameters. Therefore, these processes are not suitable for the economic large scale production of filled CNTs on the desired substrates. Furthermore, despite recent advancements in CNT synthesis, studies on the growth temperature dependence of the metal filled CNTs are still relatively rare in the literature.

Thus, from the technological point of view, it is absolutely necessary to devise an economic and scalable single stage synthesis method for the magnetic material filled CNT and also to study their growth temperature dependence.

In this article, we report a novel method for the large scale synthesis of Fe filled CNTs employing atmospheric pressure chemical vapor deposition (APCVD) of propane on Si using a modified photoresist (Mod-PR) with a metalorganic molecular precursor, Fe(acac)$_3$. The effect of growth temperature in the range of 550–950 °C on the Fe filled CNTs was also investigated. Scanning electron microscopy (SEM), X-ray diffraction (XRD), high resolu-tion transmission electron microscopy (HRTEM), energy dispersive X-ray (EDX) and Raman spectroscopy were used to characterize the morphology, phase, internal structure and quality of the resultant products.

## 2. Experimental procedure

To obtain Mod-PR solution of concentrations 0.2 M, 706.4 mg of Fe(acac)$_3$ was mixed with 10 ml of HPR 504 (Fuji Film). HPR 504 is a positive photoresist which uses ethyl lactate as the solvent and has a viscosity of 40 cps. The solution was then stirred and sonicated for 30 min to achieve a good dispersion of the metallorganic molecular precursor. After that the solution was spin-coated with a rotation speed of 4000 rpm for 20 s on the Si(1 1 1) substrate to get a thin layer of the Mod-PR. The thin Mod-PR film was then annealed in air for 10 min at 200 °C to improve the adhesion to the substrate.

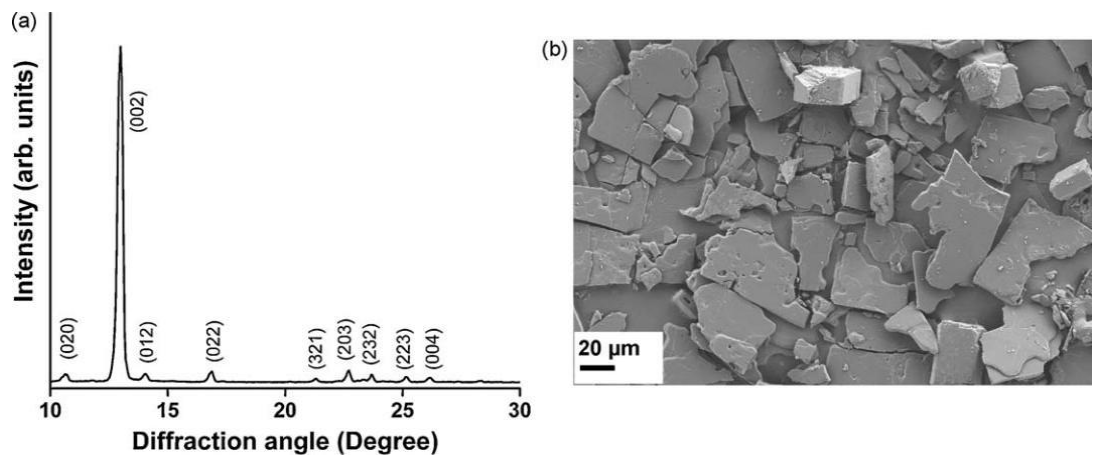

Fig. 1. (a) XRD spectrum and (b) SEM image of the precursor.

Annealed substrates were loaded into a quartz tube furnace, pumped down to 10⁻² Torr and backfilled with flowing argon to atmospheric pressure. The samples were then heated in argon up to the growth temperature following which the argon was replaced with hydrogen. Subsequently, the samples were annealed in hydrogen atmosphere for 10 min. Finally, the hydrogen was turned off; thereafter propane was introduced into the gas stream at a flow rate of 200 sccm for 1 h for CNT synthesis. The synthesis of Fe filled CNTs was performed at 550, 650, 850 and 950 °C.

SEM (ZEISS SUPRA 40) and HRTEM (JEOL JEM 2100) equipped with EDX (OXFORD Instruments) were employed for examination of the morphology and microstructure of the products. Samples were also characterized by a Philips X-ray diffractometer (PW1729) with a Cu source and a $\theta$–$2\theta$ geometry to analyze the crystallinity and phases of the products. Raman measurements were carried out with a RENISHAW RM1000B LRM at room temperature in the backscattering geometry using a 514.5 nm air-cooled Ar⁺ laser as an excitation source for compositional analysis.

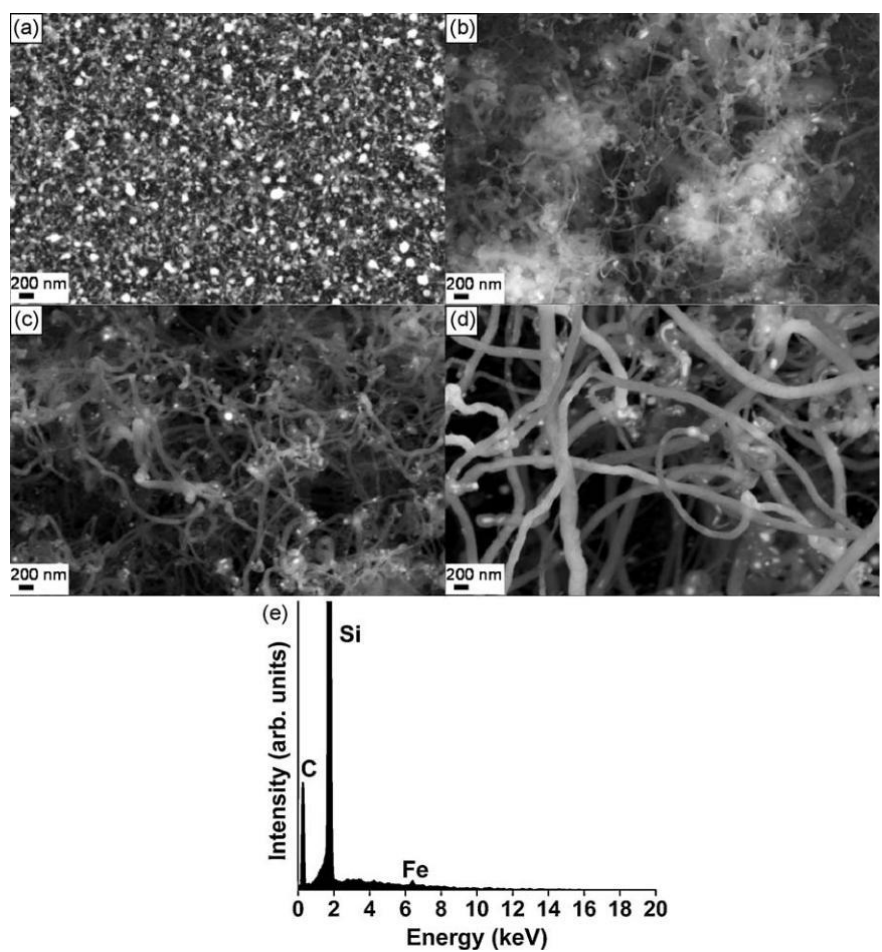

Fig. 2. SEM images of carbon nanotubes grown using Mod-PR at different temperatures (a) 550 °C, (b) 650 °C, (c) 850 °C, (d) 950 °C and (e) EDX spectrum obtained from the sample grown at 850 °C.

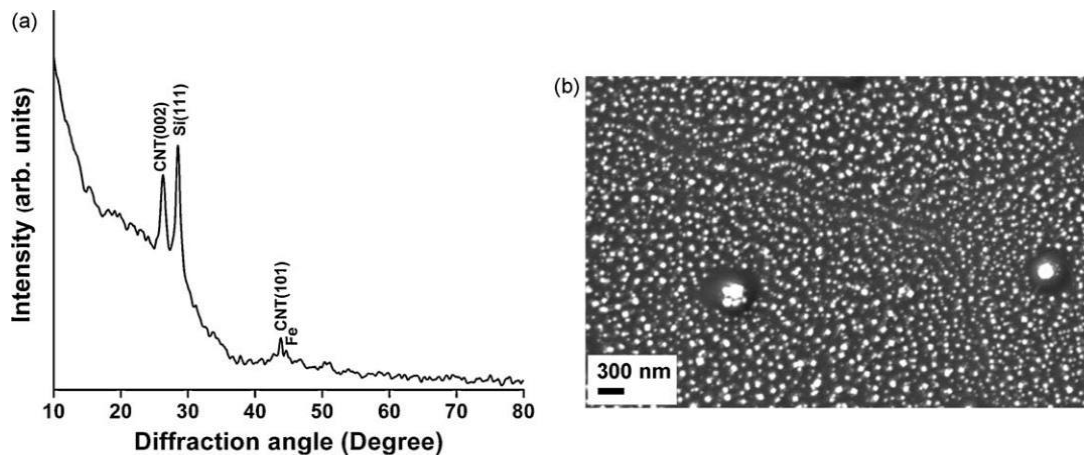

Fig. 3. (a) The XRD spectrum of the MWCNTs prepared from 0.2 M Mod-PR solution after CVD growth at 850 °C and (b) SEM micrograph of the catalytic nanoparticles prepared from 0.2 M Mod-PR solution after annealing at 850 °C.

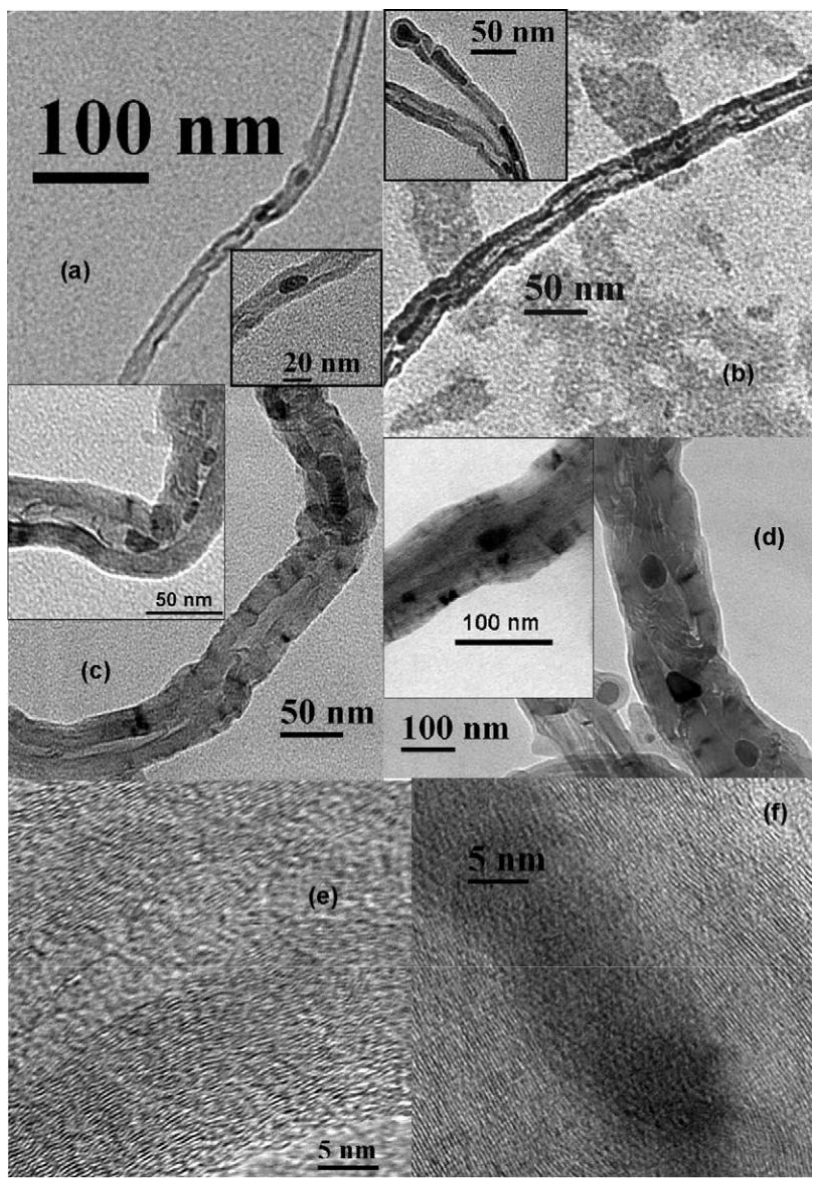

Fig. 4. HRTEM images of carbon nanotubes grown at different temperatures: (a) 550 °C, (b) 650 °C, (c) 850 °C, (d) 950 °C, (e) lattice image from a CNT grown at 650 °C and (f) lattice image from a CNT grown at 850 °C.

## 3. Results and discussion

XRD measurements were carried out to investigate the structure of the precursor and the resulting θ–2θ scan is shown in Fig. 1a. All its diffraction peaks can be assigned to the orthorhombic structure of Fe(acac)$_3$ which is in agreement with the JCPDS card no. 30-1763. SEM image (Fig. 1b) shows that the precursor comprises thin platelets with a wide range of crystallite sizes.

The analysis of the morphology and number density of the as-grown CNTs was performed using SEM. Fig. 2a–d are the SEM images of nanotubes synthesized at 550, 650, 850 and 950 °C on the Si(1 1 1) substrate using Mod-PR. The micrographs reveal that the synthesized CNTs are randomly oriented and curved with a high aspect ratio. The observation of Fe nanoparticles at the tip of the CNTs supports the tip-growth mechanism. SEM images suggest that the diameter of CNTs increases with the increase of growth temperature whereas the number density of the CNTs decreases. The surface diffusion of iron nanoparticles over Si substrate is responsible for the change of morphology of the CNTs. With the increase in the growth temperature, the surface diffusion of iron nanoparticles on the Si surface increases and significant agglom-eration of iron nanoparticles occur. Therefore at higher tempera-tures during the thermal CVD process, the size of the iron nanoparticles increases but their number density decreases [18], which eventually determines the diameter and number density of the grown nanotubes. EDX of the CNTs (Fig. 2e) shows that the material contains only carbon and Fe with a Si peak (due to substrate).

XRD measurement was performed using Cu Kα radiation (λ = 1.54059 Å) on the CNT samples grown at 850 °C and the resulting θ–2θ scan is shown in Fig. 3a. The characteristic graphitic peaks at 26.2° and 43.7° signify the presence of multi-walled carbon nanotubes (MWCNTs) in the sample. The peak near 44.7° is assigned to α-Fe (JCPDS 06-0696). The peak near 28.5° is attributed to the (1 1 1) plane of the Si substrate (JCPDS 27-1402). Fig. 3b shows the SEM image of the Fe catalytic nanoparticles formed on the Si substrate after annealing the Mod-PR film at 850 °C.

The growth morphology and internal structure of the CNTs were further investigated using high resolution TEM. Fig. 4a–d are the HRTEM images of CNTs synthesized at 550, 650, 850 and 950 °C respectively, and illustrate that the nanotubes produced at all four temperatures are multi-walled. The micrographs reveal that the diameter of the CNTs progressively increased with the increase of growth temperature. Further observations show that the CNTs are partially filled with metal nanoparticles and no nanoparticles are situated on the outer surface of the nanotubes. It indicates that our method is a simple and effective one to introduce metal inside the CNT. It can be noted that the diameter of the nanoparticles encapsulated by the CNTs is determined by the diameter of the nanotube cavity. TEM images also reveal that with the increase of growth temperature, the internal structure of the CNTs becomes more complex. In case of the CNTs grown at 550 and 650 °C, the clear cavity can be viewed. At 850 °C the CNT wall thickness is so large that in some portions the cavity tends to collapse where the walls are very close to each other. Furthermore, in the CNTs grown at 950 °C, it is very difficult to identify the cavity clearly.

To inquire into the effect of the growth temperature on the crystallinity of the nanotubes, high resolution images of the graphitic sheets of CNTs were obtained. The HRTEM image of a CNT grown at 650 °C (Fig. 4e) shows the wavy structure of graphitic sheets at short range. This wavy structure is created due to defects present in the graphite sheet. In contrast, the CNT grown at 850 °C (Fig. 4f) has well-ordered and straight lattice fringes. The dark portion between the CNT walls (Fig. 4f) indicates the presence of a metal nanoparticle. HRTEM observations suggest that the crystal-linity of CNTs improves as the growth temperature increases.

Extensive chemical composition analysis of the elongated nanoparticles, encapsulated by CNTs grown at different tempera-tures was performed. The investigation confirms that in all the cases, the encapsulated particle consists only of Fe (Fig. 5). No other phases of Fe were observed. The Cu signals, present in all the spectra, come from the Cu grid itself. By observing the results of the above mentioned analyses, the growth model of partially Fe filled CNT can be explained by emphasizing the role of the capillary action of liquid-like iron particles that exist at the time of nanotube nucleation. The detail of the growth model has been discussed elsewhere [19].

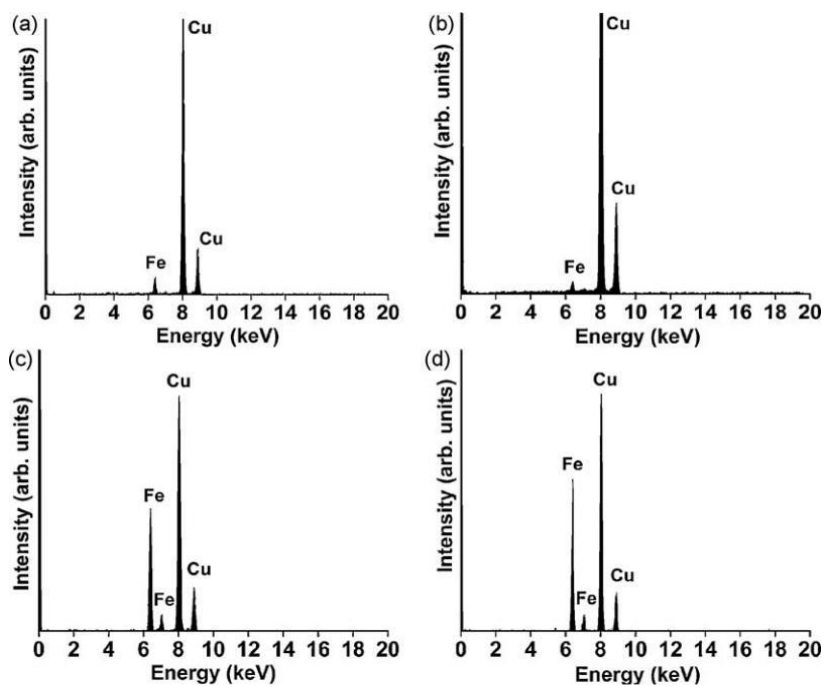

Fig. 5. The EDX spectra of the CNT encapsulated nanoparticles from the samples grown at (a) 550 °C, (b) 650 °C, (c) 850 °C and (d) 950 °C.

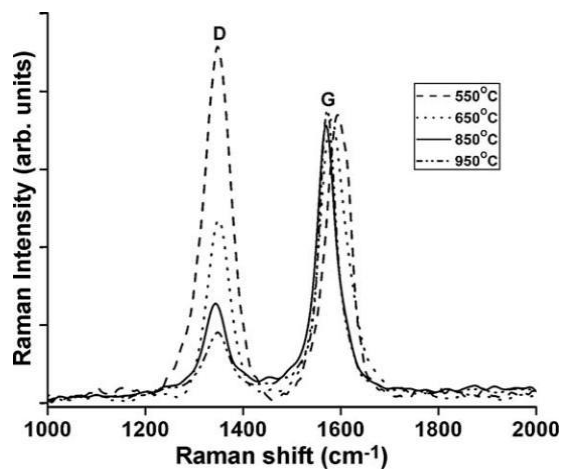

Fig. 6. Raman spectra (514.5 nm excitation) of MWCNTs grown at different temperatures.

Raman spectroscopy is also a very effective and informative method for characterization of CNTs. Fig. 6 shows the room temperature Raman spectra of the MWCNT material at a laser excitation wavelength of 514.5 nm. The two main peaks of the spectra are labeled as the D and G bands. The G band is assigned to the tangential stretching mode of highly ordered crystalline graphite and the D band represents the defects in the graphitic sheets of CNTs. The strength of the D band relative to the G band is a measure of the amount of disorder in the CNTs [20]. The values of ($I_D/I_G$) for the CNT synthesized at 550, 650, 850, and 950 °C are 1.23, 0.64, 0.35 and 0.24 respectively. The highest value of $I_D/I_G$, obtained for CNTs grown at 550 °C, suggests that the tubes grown at this temperature are more defective as compared to those grown at higher temperatures. Above 550 °C, the $I_D/I_G$ ratio decreases with increasing growth temperature, reflecting an increasing degree of graphitization in the material.

## 4. Conclusions

A novel and scalable method has been demonstrated for the production of Fe filled CNTs using a mixture of Fe(acac)$_3$ with conventional photoresist followed by CVD growth. The study reveals that the average diameter of CNTs increases with growth temperature from 550 °C to 950 °C, whereas their number density decreases. Moreover, the crystallinity of the CNTs improves progressively with increasing growth temperature. These results together demonstrate that the number density, diameter and crystallinity of partially Fe filled MWCNTs can be effectively controlled and optimized by adjusting the growth temperature during the large scale production.

**Acknowledgements**

We are grateful to Dr. B. Mishra from the department of Geology
& Geophysics, IIT Kharagpur for his help with the Raman measurement. J. Sengupta is thankful to CSIR (India) for providing a senior research fellowship.